\documentstyle[preprint,aps,epsf]{revtex}
\begin{document}

\draft

\title{Propagation of nonlinear waves in disordered media}

\author{B. Spivak}

\address{Physics Department, University of Washington, Seattle, WA 98195,
USA}

\author{A. Zyuzin}

\address{A.F.Ioffe Institute, 194021 St.Petersburg, Russia}

\maketitle

\begin{abstract}
We study the propagation of stationary waves in disordered
non-linear
media
 described by the nonlinear
Schr\"odinger equation and show that for given boundary conditions and a 
given coherent wave incident on the sample,
the number of solutions of the equation increases exponentially
with sample size. We also discuss the ballistic case, 
the sensitivity of the solutions to the change of external parameters, the
similarity of this problem to the problem of
spin glasses and time-dependent solutions.
\end{abstract}

\section{Introduction}
In this article we consider the propagation of a stationary coherent wave
described by
a field $\phi({\bf r})$, in a nonlinear elastically
scattering medium.
Though we believe that our results are of a general character, for the
sake
of concreteness we consider
the situation where the wave is described by a
nonlinear Schr\"odinger equation
\begin{equation}
\{-\frac{1}{2m}\frac{\partial^2}{\partial {\bf r}^2}-\epsilon+u({\bf r})+
\beta n({\bf r})\}\phi({\bf r})=0
\end{equation}
Here $n({\bf r})=|\phi({\bf r})|^{2}$ is the density,  $m$ is
the wave mass, $\epsilon$ is the wave energy, $\beta$ is a constant and
 $u({\bf r})$ is a
scattering potential, which is a random function of the coordinates.
Eq.1 appears, for example, in the theory of electromagnetic waves
propagating in nonlinear media \cite{Landau}, in the theory of hydrodynamic
turbulence \cite{Zakharov}, and in the theory of turbulent plasma
\cite{Kadomzev}.

We will assume that $u({\bf r})$ obays the white noise statistics:
 $\langle u({\bf r})\rangle=0$ ,
$\langle u({\bf r})u({\bf r_{1}})\rangle=
\frac{\pi}{lm^{2}}\delta({\bf r}- {\bf r}_{1})$.
Here brackets $\langle \rangle$ correspond to averaging over realizations
of $u({\bf r})$, and  $l$ is the elastic
mean free path ($l\gg k^{-1}=(2\epsilon m)^{-\frac{1}{2}}$).
In the presence of the scattering potential $u({\bf r})$, the
 spatial dependence of
the $n({\bf r})$ exhibits random sample-specific fluctuations, which are
called speckles.  In the case of elastically 
scattering diffusive linear media this problem was considered in
\cite{ZuyzinSpivak,Kane,ZyuzinSpivakRev}. 
Below we will be interested in the statistics of $n({\bf r})$ in the case
of
nonlinear diffusive media. In particular, we
 will show that, for given boundary conditions,
the number of solutions of Eq.1  increases exponentially with sample
size. A brief summary of some of the results has been published in
\cite{SpivakZyuzinNL}.

\section{Nonlinear speckles.}
 
Let us consider the case where a coherent wave
$\phi_{0}({\bf r})=\sqrt{n_{0}}e^{i{\bf kr}}$
with momentum ${\bf k}$ is incident on a disordered sample of
size $L\gg l$ (See the insert in Fig.1a).

 The ${\bf r}$-dependence of the average
density
$\langle n({\bf r})\rangle$
can be
described by the
diffusion
equation,
\begin{eqnarray}
\langle {\bf j}({\bf r})\rangle=-D\frac{\partial \langle n({\bf
r}) \rangle}{\partial
{\bf r}} \nonumber \\
div \langle {\bf j}({\bf r}) \rangle=0
\end{eqnarray}
 which is equivalent the to calculation of the diagrams shown in
Fig.2.a.
Here $ {\bf j}({\bf r})$ is the current density. In the limit $k^{-1}\ll l\ll L$ the expression for the diffusion 
coefficient $D=\frac{lk}{3m}$ has a classical form.
In the case of the geometry shown in Fig.1, the total flux 
through the sample is zero, and the average density
 $\langle n({\bf r}) \rangle=n_{0}$ is spatially uniform.

The term $\beta |\phi({\bf r})|^{2}$ in Eq.1  plays the role of an
additional 
scattering potential. It will be shown that its contribution 
to the diffusion coefficient $D$ can be neglected
 at small enough intensities of the incident beam,
when  
\begin{equation}
|\beta n_{0}|^{2} \ll
\frac{\epsilon k}{lm}.
\end{equation}
At $\beta >0$ the propagation of a uniform wave in an infinite medium
described by Eq.1 is unstable at arbitrarily small $n_{0}$ due to the
effect of nonlinear self
focusing. The characteristic length at which the self-focusing
takes place is of order $r_{(sf)}\sim \epsilon k/(\beta n_{0})^{2}m$
 \cite{Landau}. Thus the inequality Eq.3 is equivalent to $r_{(sf)}\gg l$.
One can say that the regime Eq.3 corresponds to a system of randomly 
distributed weak concave and convex lenses. 

The diffusion equation approximation completely neglects
interference effects, which lead to the existence of speckles.
To describe them one should solve Eq.1 before averaging over
the realizations of $u({\bf r})$.
It is convenient to            
expand the density 
\begin{equation}
\beta n({\bf r})=\frac{D}{\sqrt{L}}\sum_{i=1}^{\infty} 
i^{1/3}\bar{u}_{i}n_{i}({\bf r})
\end{equation}
in a complete set
of orthonormal eigenstates $n_{i}({\bf r})$ 
of the diffusion equation $(\int d {\bf r} n_{i}^{2}({\bf r})=1)$:
\begin{equation}
D\frac{\partial^{2}}{\partial^{2} {\bf r}}n_{i}({\bf r})=E_{i}n_{i}({\bf
r})
\end{equation}
where $E_{i}\sim \frac{D}{L^{2}}i^{2/3}$ are the eigenvalues  
of Eq.5, and $i=1,2...$ labels the eigenstates.
We will show below why the expansion Eq.4 is convenient. We assume
boundary
conditions corresponding to zero current through a closed boundary,  
and $n({\bf r})=0$ at the open boundary.

Let us first substitute Eq.4 into the nonlinear term of Eq.1,
 and regard the $\{\bar{u}_{i}\}=\bar{u}_{1},... \bar{u}_{j},...$ as
 independent parameters. This gives a linear equation for $\phi({\bf r})$
\begin{equation}
\left(-\frac{1}{2m}\frac{\partial^2}{\partial {\bf r}^2}-\epsilon+u({\bf
r})+
\beta\frac{D}{\sqrt{L}}\sum_{i=1}^{\infty}
i^{1/3}\bar{u}_{i}n_{i}({\bf r})\right)\phi({\bf r})=0
\end{equation}
Denoting a solution of Eq.6 at a given set of
parameters $\{\bar{u}_{i}\}$, as
$\phi({\bf r}, \{ u({\bf r})\}, \{\bar{u}_{m}\})$ and constructing 
$n({\bf r}, \{ u({\bf r}\}), \{\bar{u}_{i}\})=
|\phi({\bf r}, \{u({\bf r})\}, \{\bar{u}_{i}\})|^{2}$, we can
write the
self-consistency equations for $\{\bar{u}_{i}\}$:
\begin{equation}
\frac{i^{2/3}\bar{u}_{i}}{\gamma}=F_{i}
(\{\bar{u}_{i}\}).
\end{equation}
Here
\begin{equation}
\gamma =|\frac{3n_{0}\beta}{2\epsilon}|(\frac{L}{l})^{3/2}
\end{equation}
and
\begin{equation}
F_{i}(\{\bar{u}_{i}\})=\frac{ki^{1/3}l^{1/2}}{n_{0}L}\int
d {\bf r} n({\bf r},\{u({\bf r})\},
\{\bar{u}_{i}\}) n_{i}({\bf r})
\end{equation}
 are
dimensionless random functions of $\{\bar{u}_{i}\}$, the form of which 
depends on $u({\bf r})$.

 To investigate the properties of the solutions of
Eq.7,
we have to know the statistical properties of the random functions $F_{i}
(\{\bar{u}_{i}\})$. We can infer these properties from 
values of different correlation functions of 
$F_{i}
(\{\bar{u}_{i}\})$ obtained by averaging
 over realizations of  $u({\bf
r})$. 
It is important that the statistical analysis of the
random functions $F_{i}(\{\bar{u}_{i} \})$ is
equivalent to the analysis of linear speckles, which has been done in
 \cite{ZuyzinSpivak,Kane,ZyuzinSpivakRev}.
To characterize  $F_{i}(\{\bar{u}_{i}\})$ we  calculate the
following
correlation functions 
\begin{equation}
\langle \delta F_{i}(\{\bar{u}_{i}\})\delta
F_{j}(\{\bar{u}_{i}\})\rangle=\delta_{ij},
\end{equation}
\begin{equation}
\langle [F_{i}(\{\bar{u}_{i}+\Delta
\bar{u}_{i}\})-F_{i}(\{\bar{u}_{i}\})]^{2}\rangle \sim
(\Delta\bar{u}_{n})^{2} \qquad at \qquad \Delta \bar{u}_{i}\sim 1.
\end{equation}
\begin{equation}
\langle \frac{\partial F_{i}}{\partial \bar{u}_{r}}\times \frac{
\partial F_{j}}{\partial \bar{u}_{s}}\rangle \sim 
((r/s)^{1/3}+(s/r)^{1/3})\frac{1}{(|i-j|+|r-s|)^{2/3}}
\end{equation}
Here $\delta
F_{i}=F_{i}-\langle F_{i} \rangle$ and $\langle
F_{i}(\{\bar{u}_{i}\}) \rangle =const$ which is independent of
$\{\bar{u}_{i}\}$. 
We present the derivation of Eqs.10-12 in the next section.
The simple form of Eqs.10-12 is a consequency of the choice of $n_{i}({\bf
r})$ in
 Eqs.4,5.
Eq.10 indicates that mesoscopic fluctuations of
different functions $F_{i}$ are
uncorrelated. According to Eq.12, the correlation of derivatives of
 $F_{i}(\{\bar{u}_{i}\})$  over different $\bar{u}_{r}$ is small for 
$|r-s|>1$. It will be shown
in
the
next section that these facts are 
consequences of the choice of $n_{i}({\bf r})$ in Eq.4 as
eigenfunctions of Eq.5.
The introduction of the coefficients $i^{1/3}D/\sqrt{L}$ in Eq.4 ensures
Eqs.10,11 appears as above.
 
Thus we arrive at the following picture: the $F_{i}(\{\bar{u}_{i}\})$
fluctuate 
randomly as functions of $\{\bar{u}_{i}\}$ near their average, which is
independent of $\{\bar{u}_{i}\}$. According to Eq.11, the characteristic
period of the
fluctuations
is of order one. The fluctuations both of different functions,
and the same functions with respect to different
$\bar{u}_{i}$ are uncorrelated.

Using this information about the $F_{i}(\{ \bar{u}_{i} \})$
 we can estimate the number of  solutions of
 Eq.7  (or Eq.1). If $\gamma\ll 1$, Eq.7 has a unique solution while for
$\gamma\gg 1$ Eq.7 has many solutions.
Let us consider the $i^{th}$ equation in Eq.7 and fix all variables
$\bar{u}_{j}$ other
than $\bar{u}_{i}$. Then, at $i^{2/3}\gamma^{-1}\gg 1$ the equation has
a unique solution, while at $i^{2/3}\gamma^{-1}\ll 1$ the number of the
solutions
is of order $\gamma i^{-2/3}$.
In Fig.1a we show a qualitative graphical solution of Eq.7, which
corresponds
to the
intersection of two functions: $F_{i}(....,\bar{u}_{i},....)$ and
$\gamma^{-1}i^{2/3}\bar{u}_{i}$. The solutions are distributed in an 
interval
of order $\gamma i^{-2/3}$.

Therefore, to estimate number of solutions of Eqs.7 at
$\gamma\gg 1$, we have
to take into account only a subset of Eqs.7 with $i<I=\gamma^{3/2}$.
Since both the amplitude of fluctuations and the periods in the $i^{th}$
direction of 
randomly
rippled hypersurfaces
$F_{i}(\{\bar{u}_{i}\})$ are of order unity,
the number of solutions $\textit{N}$ of Eqs.1,7  is
proportional to the volume of an $I$-dimensional hyperparallelepiped
with sides of order 
$\gamma i^{-2/3}$,
$i<I$. As a result we have
\begin{equation}
\textit{N} \sim
\gamma^{I}\prod_{1}^{I}i^{-2/3}\sim \exp(\frac{2}{3}\gamma^{3/2})
\end{equation}
 Thus the number of the solutions $\textit{N}$ of Eq.1
increases
exponentially with the sample size $L$.

To illustrate Eq.13 we consider the case $I=2$ ($\gamma\sim 1$). Then
Eqs.7
can be viewed
as two surfaces $z=F_{1}(\bar{u}_{1}, \bar{u}_{2})$ and
$z=F_{2}(\bar{u}_{1},
\bar{u}_{2})$, which are intersected by two planes
$z=\gamma^{-1}\bar{u}_{1}$ and
$z=\gamma^{-1}2^{2/3}\bar{u}_{2}$ respectively. A result of these
intersections is two systems of lines in the plane
$\bar{u}_{1},\bar{u}_{2}$ shown in Fig.1b. 
 The solid lines correspond to 
intersections between
 $z=F_{1}(\bar{u}_{1}, \bar{u}_{2})$ and $z=\gamma^{-1}\bar{u}_{1}$, 
and are located within a strip in the
$\bar{u}_{1}$ direction of width  of order $\gamma$. 
The dashed lines correspond to the intersection of the surfaces
$z=F_{2}(\bar{u}_{1},\bar{u}_{2})$ and $z=\gamma^{-1}2^{2/3}\bar{u}_{2}$,
and 
are located within a strip of width $\gamma 2^{-2/3}$ in
$\bar{u}_{2}$ direction. 
 The intersections of solid and dashed lines in Fig.1b correspond
to solutions of the
system of Eqs.7.  According to Eqs.10-12, the dashed and the solid
lines in Fig.1b are uncorrelated. The typical distance between, say,
solid
lines is of order one. As a result, the number of solutions
$N_{I=2}$ of Eqs.7 in this case is of the order of the area of
a parallelogram with sides 
$\gamma$ and $2^{-2/3}\gamma$
\begin{equation}
N_{I=2}\sim \gamma\times 2^{-2/3}\gamma.
\end{equation}   

Now let us estimate the corrections to the diffusion coefficients
 originating from  scattering from the 
potential $\beta n({\bf r})$.
In the case $|{\bf r-r'}|<l$ we have
($D=3$) \cite{ZyuzinSpivakRev}
\begin{equation}
\langle n({\bf r})n({\bf r}')\rangle =
\frac{n_{0}^{2}m^{2}}{ k^{2} |{\bf r-r'}|^{2}},
\end{equation}
 independently
of the values of the parameters $\{\bar{u}_{i}\}$.
 In the Born approximation, the nonlinear mean free path
corresponding to scattering from the potential $\beta n({\bf r})$ is:

\begin{equation}
l_{(nl)}=\frac{\epsilon k^{2}}{(\beta n_{0})^{2} m}
\end{equation}
Thus the criterion Eq.3 is equivalent to $l_{(nl)}=r_{(sf)}\gg l$.

In the opposite limit $l_{(nl)}\ll l$, the scattering mean free path
 is determined by the 
scattering from the potential $\beta n({\bf r})$, and one can
neglect the random potential $u({\bf r})$ in Eq.1. In this case one
can estimate the number of solutions of Eq.1 by substituting
 $l_{(nl)}$
instead of $l$ into Eqs.8, or by substituting $\gamma_{(nl)}$ into Eq.13
instead of
$\gamma$, where 
\begin{equation}
\gamma_{(nl)}=\frac{3}{2}\frac{(n_{0}\beta)^{4}}{\epsilon}(\frac{Lm}{\epsilon
k})^{3/2}
\end{equation}

In the conclusion of this section we would like to discuss 
the condition $\gamma >1$ for the existence of multiple solutions of
Eq.1.
In the absence of the nonlinear term $\beta n({\bf r})$ the solution of
Eq.1 corresponds 
to particles traveling along diffusion trajectories.
In the presence of the nonlinear term
 $\beta n({\bf r})$, the probability amplitude for
traveling along a diffusive trajectory
acquires an additional phase of order $\delta \chi
_{(nl)}=\frac{\beta k}{2}\int ds\delta n({\bf r})\sim 1$, 
where the integration is taken along a diffusive trajectory.
To estimate the value of the additional phase let us calculate the integral
\begin{equation}
\left\langle \left( \delta \chi _{(nl)}\right) ^2\right\rangle =\left(
\frac{\beta k}2\right) ^2\int dsds^{\prime }\left\langle \delta n({\bf r}
)\delta n({\bf r}^{\prime })\right\rangle
\end{equation}
The correlation function of densities is given by Eq.15 and
the estimate for Eq.18 is $(\beta k)^{2}\frac{L^{3}}{l^{2}}$.
Thus the criterion $\gamma>1$ corresponds to  $\langle( \delta
\chi_{(nl)})
^{2}\rangle >1 $.
Another interpretation is that at $\gamma>1$
the
 sensitivity of the solutions of Eq.1 to a change of $u({\bf r})$ 
increases significantly as compared
to its single particle value \cite{SpivakZyuzinNL}.

\section{Calculation of correlation functions of Eqs.11-13}

Now we turn to the calculation of correlation functions Eqs.10-12.
According to the definition Eq.9 we get 
\begin{equation}
\langle F_{i}(\{\bar{u}_{i}\}) F_{j}(\{\bar{u}_{i}+\Delta
\bar{u}_{i}\})
\rangle=\frac{i^{1/3}j^{1/3}lk^{2}}{L^{2}n_{0}^{2}}
\int d {\bf r} d {\bf r}' n_{i}({\bf r}) n_{j}({\bf r}') 
\langle n({\bf r}, \{\bar{u}_{l}\})n({\bf r}_{1},
\{\bar{u}_{l}+\Delta \bar{u}_{l} \})\rangle
\end{equation}
In the approximation Eq.3, 
 the value of $\langle n({\bf r}, \{\bar{u}_{l}\})n({\bf r}_{1},
\{\bar{u}_{l}+\Delta \bar{u}_{l} \})\rangle$ is
 independent of $\{\bar{u}_{l}\}$ and has the same form as in the
linear case \cite{ZuyzinSpivak,ZyuzinSpivakRev}. Therefore,
 one can use the standard diagram technique for
averaging
over $u({\bf r})$ \cite{Abrikosov}. 
The diagrams describing the correlation functions
$\langle n({\bf r},\{\bar{u}_{l}\}) n({\bf
r}',\{\bar{u}_{l}+\Delta \bar{u}_{l} \})\rangle$
are shown in
Fig.2b,c. 
Alternatively, one can solve the Langevin equation
 \cite{ZuyzinSpivak,ZyuzinSpivakRev}, valid at
$|{\bf r-r'}|\gg l$ 
\begin{equation}
div \delta {\bf j}({\bf r})=0
\end{equation}
\begin{equation}
\delta {\bf j}({\bf r})=-D\frac{\partial}{\partial {\bf r}} \delta n({\bf
r})+{\bf J}^{L}({\bf
r}, \{u({\bf r})\}, \{\bar{u}_{i}\}).
\end{equation}
The correlation function of random Langevin forces ${\bf J}^{L}({\bf
r})$ is given by the diagram shown in Fig.2b. It can be written as
\begin{eqnarray}
\langle J_{i}^{L}({\bf r}, \{u({\bf r}\},\{\bar{u}_{i}\})J_{j}^{L}({\bf
r'}, 
\{u({\bf r})\}, \{\bar{u}_{i}+\Delta \bar{u}_{i}\})\rangle = \nonumber
\\
\frac{2\pi l}{3m^{2}}
 |\langle \phi({\bf r},\{\bar{u}_{l} \})\phi^{*}({\bf r}, \{
\bar{u}_{l}+\Delta \bar{u}_{l} \}) \rangle |^{2}\delta_{ij}
\delta({\bf r-r'})
\end{eqnarray}
The expression for $\langle \phi({\bf r},\{\bar{u}_{l} \})\phi^{*}({\bf
r}, \{
\bar{u}_{l}+\Delta \bar{u}_{l} \}) \rangle$ is given by the ladder
diagrams shown in Fig.2b,
where the inner Green function
correspond to $\{\bar{u}_{i}\}$, while the outer Green functions
correspond to $\{\bar{u}_{m}+\Delta \bar{u}_{m}\}$.
Eq.22 is a generalization of corresponding equations introduced in
\cite{ZuyzinSpivak,ZyuzinSpivakRev}. Namely, in the case $\Delta
\bar{u}_{l}=0$ one has to substitute
$\langle n \rangle^{2}=n_{0}^{2}$ in Eq.22  instead of $|\langle \phi({\bf
r},\{\bar{u}_{l} \})\phi^{*}({\bf r}, \{
\bar{u}_{l}+\Delta \bar{u}_{l} \}) \rangle |^{2}$.

Using Eqs.20-22 we get
\begin{eqnarray}
\langle n({\bf r}, \{ \bar{u}_{l} \}) n({\bf r}',
\{\bar{u}_{l}+\Delta \bar{u}_{l} \}) \rangle= \nonumber \\
\frac{2 \pi l}{3 m^{2}}
\int d {\bf r}_{1} \frac{d \Pi({\bf r}, {\bf r}_{1})}{d {\bf r}_{1}} 
\frac{d \Pi({\bf r}',{\bf r}_{1})}{d {\bf r}_{1}} 
|\langle \phi({\bf r},\{\bar{u}_{l} \})\phi^{*}({\bf r}, \{
\bar{u}_{l}+\Delta \bar{u}_{l} \}) \rangle |^{2}.
\end{eqnarray}
\begin{equation}
\Pi({\bf r}, {\bf r}')=\sum_{l} \frac{n_{l}({\bf r}) n_{l}({\bf
r}')}{E_{l}}
\end{equation}
Here $\Pi({\bf r}, {\bf r}')$ is the Green's function of Eq.5.

At $\Delta \bar{u}_{i}=0$, doing the integral over ${\bf r}_{1}$ in Eq.23
by parts, using Eq.22
and taking into account the orthogonality of the functions $n_{i}({\bf
r})$,  we get Eq.10.
The diagrams in Fig.2.b for  $\langle \phi({\bf r},\{\bar{u}_{l}
\})\phi^{*}({\bf r}, \{
\bar{u}_{l}+\Delta \bar{u}_{l} \}) \rangle$ are equivalent to 
the equation \cite{AltshulerSpivak,FengLeeStone}
\begin{equation}
\left(D\frac{\partial^{2}}{\partial^{2} {\bf
r}}+i\frac{D}{\sqrt{L}}\sum_{i}i^{1/3}\Delta \bar{u}_{i}n_{i}({\bf
r})\right) 
\langle \phi({\bf r},\{\bar{u}_{l} \})\phi^{*}({\bf r}, \{
\bar{u}_{l}+\Delta \bar{u}_{l} \}) \rangle=0
\end{equation}
The existence of the term proportional to $\Delta \bar{u}_{i}$ in Eq.25 
reflects the fact that the inner lines in the diagrams shown in Fig.2b
correspond to the Green's functions in the scattering potential
characterized by $\{\bar{u}_{i}\}$ while the outer lines correspond to 
those of $\{\bar{u}_{i}+\Delta \bar{u}_{i}\}$. 
Note that as long as Eq.3 holds Eq.25 is independent of
$\{\bar{u}_{i}\}$.

Solving Eq.25 by perturbation theory with respect to 
$\Delta \bar{u}_{l}$ we get  
\begin{eqnarray}
\langle \phi({\bf r},\{\bar{u}_{l} \})\phi^{*}({\bf r}, \{
\bar{u}_{l}+\Delta \bar{u}_{l} \}) \rangle= \nonumber \\
n_{0}\left( 1+i\frac{D}{\sqrt{L}}\sum_{i}i^{1/3}\frac{n_{i}({\bf
r})}{E_{i}}
\Delta \bar{u}_{i}-\frac{D^{2}}{L} \int d {\bf r}' \Pi({\bf r},
{\bf r}')\sum_{i} \frac{i^{2/3}(\Delta \bar{u}_{i}n_{i}({\bf
r}'))^{2}}{E_{i}^{2}}+ ...\right).
\end{eqnarray} 

We can neglect the second term in brackets in Eq.26 because it is of order
$(i^{-1/3}\Delta \bar{u}_{i})$ (and its contribution to Eq.22 is of order
$i^{-2/3}(\Delta \bar{u}_{i})^{2}$), while the contribution to Eq.22 from
the
third term is $(\Delta \bar{u}_{i})^{2}$.
To get the latter estimate we took into account that $\Pi({\bf r,r'})\sim 
(DL)^{-1}$ at $|{\bf r-r'}|\sim L$. Substituting Eq.26 into Eq.19,22,23,
we get Eqs.11,12.

 \section{Discussion}

 The estimate Eq.14 was made for the case of
a typical realization of the scattering potential. On the other hand, even
at $\gamma<1$, there are rare realizations of $u({\bf r})$, which
correspond to several solutions of Eq.1.

The results presented above hold
for arbitrary sign of $\beta$. This is quite different from the situation
in the pure case ($u({\bf r})=0$) \cite{Landau} where at $\beta >0$ 
self-focusing
takes place.

At $\gamma\gg 1$, the solutions of Eq.1 exhibit 
exponentially large
sensitivity to changes of parameters of the system
\cite{SpivakZyuzinNL}. Consider, for
example,
the case where the incident angle $\theta$ of the
wave is changing, and suppose that a solution of Eq.1
follows
this
change adiabatically. Then an exponentially small change
\begin{equation}
\Delta \theta\sim
\exp(-\frac{2}{3}\gamma^{3/2})
\end{equation}
will lead to disappearance of the solution, and the system will
 jump to another solution.

Similar phenomenon may occur
in the system of interacting electrons is
disordered metals: it can
be unstable with respect to the creation  of random
magnetic
moments. 
This would correspond to Finkelshtain's scenario
\cite{finkelshtein}. However,
in this case in order to get a self-consistency equation, which would be
an analog
of Eq.7, we have to integrate
over electron energies, which decreases the
amplitude of mesoscopic fluctuations. As a result, the situation with many
solutions may occur only in the D=2 case and the characteristic
spatial scale will be of the order of the
electron localization
length in the linear problem. Thus the problem of interacting
electrons in disordered metals remains unsolved.

We would like to mention a similarity of the problem  considered
above to the problem of spin glasses. To illustrate this point
let us consider a model in which the coefficient $\beta({\bf r})$ in Eq.1 
is nonzero only at points ${\bf r}={\bf r}_{\alpha}$, $\alpha=1,2.....$ 
\begin{equation}
\beta({\bf r})=\beta_{0}\sum_{\alpha}U({\bf r}-{\bf r}_{\alpha}),
\end{equation}
where $U(|{\bf r}|)$ is a short range function decaying on characteristic
distance $R<1/k$, and
of a maximum height $U_{0}$ ,
and ${\bf r}_{i}$ are randomly distributed in space with given density.
Then Eq.1 can be rewritten only in terms of values of $\phi_{\alpha}=
\phi({\bf r}={\bf r}_{\alpha})$
\begin{equation}
\phi_{\alpha}=\beta_{0}U_{0}R^{3}\sum_{\beta}G({\bf
r}_{\alpha}, {\bf r}_{\beta})|\phi_{\beta}|^{2}\phi_{\beta},
\end{equation}
\begin{equation}
G({\bf r}_{\alpha},{\bf r}_{\beta})\sim \frac{\exp(ik|{\bf
r}_{\alpha}-{\bf
r}_{\beta}|+i\delta ({\bf r}_{\alpha},{\bf r}_{\beta})}{|{\bf
r}_{\alpha}-{\bf r}_{\beta}|^{1/2}} \qquad |{\bf
r}_{\alpha}-{\bf r}_{\beta}|\gg l
\end{equation} 
where $G({\bf r}_{\alpha},{\bf r}_{\beta})$ is the Green function of the
linear Schr\"odinger equation, and the phase
$\delta({\bf r}_{\alpha}, {\bf r}_{\beta})$ is a random quantity at $|{\bf
r}_{\alpha}-{\bf
r}_{\beta}|\gg l$. The major difference between Eq.30 and the spin glass
problem is 
that in the former case one is interested in the minimum of the free 
energy, while in the case of Eq.30 the boundary conditions are given.
Thus there are no thermodynamic criteria on how to choose between multiple 
solutions of the stationary Eq.1. One of the possibilities is that the
state of the system 
is determined by its history. 

We may question how many of these stationary states are stable. 
 The time-dependent non-linear Schr\"odinger equation
may be obtained from Eq.1 by
substituting $\epsilon$ for $i\partial_{t}$.
Equivalently, we can write equations 
for $\bar{u}_{i}(t)$. We still assume the same stationary boundary
conditions. In the absence of a complete solution we present a
qualitative 
picture.
  
The characteristic time of change of the $i'th$ harmonic $\bar{u}_{i}$ is 
$\tau_{i}\sim E^{-1}_{i}$. Thus we have
\begin{equation}
\tau_{i}\left( 1+\gamma g_{i}(\{\bar{u}_{i}\})\right) \partial_{t}
\bar{u}_{i}(t)
= 
 \gamma i^{-\frac{2}{3}}F(\{\bar{u}_{i}\})-\bar{u}_{i}
\end{equation}
Here $g_{i}(\{\bar{u}_{i}\})\sim 1$ is a dimensionless random function. 
 The statistical properties of the function
$g_{i}(\{\bar{u}_{i}\})$ are roughly the same as
those for $F_{i}(\{\bar{u}_{i}\})$. Namely,
the amplitude and sign randomly oscillate with a characteristic periods of
order one.
Strictly speaking, Eq.31 holds if $|\tau_{i}\partial_{t}\bar{u}_{i}|\ll
1$ and the characteristic time of establishing stationary distributions
$n({\bf r})$ at given $\bar{u}_{i}(t)$ is much shorter than the
characteristic
time of change of $\bar{u}_{i}(t)$. In other words, Eqs.31 represent a
sort 
of hydrodynamics.
This takes place, for example, near the 
critical instability points (see, for example,
the
point $"a"$ on Fig1.a.). 
 Generally, $\tau_{i} \partial_{t} \bar{u}_{i}(t)\sim 1 $,
so
 we have to keep not only the first but also several
 higher time derivatives in Eqs.31. 
 We believe, however, that the model Eqs.31 captures
the
 qualitative features of the system's dynamics correctly even in this
case.
  
Linearizing Eqs.31 near the stationary solutions, we arrive at the
 conclusion that the fraction of solutions of Eq.7 which are stable is
of order $2^{-I}$. 
Thus, at $\gamma\gg 1$ the number of stable stationary
 solutions  is still exponentially large.

In principle, Eqs.31 can also have nonstationary
solutions as $t\rightarrow \infty$. Obviously, the 
characteristic amplitudes of the solutions are given by Eq.7, 
($|\delta \bar{u}_{i}(t)|<\gamma i^{-2/3}$). A complete investigation 
of  $\{\bar{u}_{i}(t)\}$ is a
complicated and still
unsolved problem.
 For example, for $I=1$ and $t\rightarrow \infty$
only stable stationary solutions are relevant. For $I=2$, and
$t\rightarrow \infty$,
 depending on the properties of the realizations of the random sample
specific
functions 
$F_{1}(\bar{u}_{1}, \bar{u}_{2})$ and $F_{2}(\bar{u}_{1}, \bar{u}_{2})$,
 one can, additionally, have periodic, quasiperiodic, and
chaoric in time solutions.
 If $I=3$, one can also, have strange (or stochastic) attractors
 as solutions of three differential equations.

In this respect we would like to mention
papers \cite{skipetrov1,skipetrov2,skipetrov3}, where an attempt to
 describe temporal nonlinear speckles was done. We believe that 
these results are incorrect.
To estimate the instability threshold it is
sufficient to expand the nonlinear Langevin equations in powers of
$\beta$ (see
diagrams shown in Fig.2 in \cite{SpivakZyuzinNL}, or in Fig.4
 in \cite{skipetrov2}). This is  why the
 authors of \cite{skipetrov1,skipetrov2,skipetrov3} were able to reproduce
the instability criterion $\gamma >1$ obtained in
\cite{SpivakZyuzinNL}.
Strictly speaking, this approach holds only
in the case when $\gamma\ll 1$.
For $\gamma> 1$ this approach describes the system incorrectly.
In the absence of solid mathematical procedure the authors of 
\cite{skipetrov1,skipetrov2,skipetrov3} made assumptions about the nature
of solutions of Eq.1 beyond the instability point $\gamma>1$. Namely,
they made assumptions that at $t\rightarrow \infty$ the finction 
$n({\bf r}, t)$  
must 
exhibit oscillations in time 
and they assumed some form of time correlations of $n({\bf r},
t)$. Both assumptions are incorrect. This can be seen,
for example, from the fact that their approach cannot
reproduce the existence of an exponentially big (for $\gamma\gg 1$) number
of multiple stationary solutions of Eqs.7, which are singular points of
Eqs.31.

The simplest situation where the deficiency of the approach of 
$\cite{skipetrov1,skipetrov2,skipetrov3}$ is most evident takes
place near the first instability point
 $\gamma=\gamma_{c}\sim 1$. This is schematically shown by the
dashed line in Fig.1a. The critical value $\gamma_{c}$ is sample
specific. In this case, say, the first
equation $(i=1)$ in the system of Eqs.7 
has three solutions. Two of them
are close to each other (see the point "a" in
Fig.1a.). Let us start with a
discussion of
the system dynamics near this point. 
We would like to stress that in this case 
 $(\tau_{1}\partial_{t}\bar{u}_{1}(t))/\bar{u}_{1}\ll 1$  and
our analysis is
rigorous. This is exactly the
 regime considered in \cite{skipetrov3}.  
Since $\partial_{t}\bar{u}_{i}/\bar{u}_{i}\sim E_{i}\gg
(\gamma-\gamma_{c})E_{1}$ for $i>1$, we can neglect the time
derivatives in all equations of the system
 Eqs.31, except the one with $i=1$. Moreover, since this point of
instability is
a rare one, all equations in Eqs.7 with $i>1$ have unique solutions.
Thus, the system's dynamics is described by just one
first order differential equation for $\bar{u}_{1}$. Near the point $"a"$
in Fig.1a
the functions
$g_{i}(\{\bar{u}_{i}\})$ and
$F_{i>1}(\{\bar{u}_{i}\})$ change slowly. Thus,
one  stationary solution is stable and the other is unstable.
The solution indicated by the letter "b" in Fig.1a is also
stable because it is related adiabatically to the unique stable solution
for $\gamma<\gamma_{c}$. At $t\rightarrow \infty$ the
system
approaches
one of the stable stationary solutions, described by Eqs.7 
\cite{SpivakZyuzinNL}. Thus the assumption made in \cite{skipetrov3} about
existence of oscillating in time solutions near the critical point is wrong. 

Far from the critical point, when $I\gg 1$ and the number of relevant
equations of the system
of Eqs.31 is larger than one,  depending on initial conditions and the
form of
random functions $F_{i}(\{\bar{u}_{i}\}$, Eqs.31 can have
periodic, chaotic solutions and strange
attractors, in addition to stationary points.
The fractions of the phase space which at $t\rightarrow
\infty$ are attracted to these 
types
of motion are currently
not known.
In any case, the solutions of the time-dependent
nonlinear Schr\"odinger equation, or Eqs.31, have a very complicated
non-Gaussian character, which is very different from that assumed in
 \cite{skipetrov1,skipetrov2,skipetrov3}. 

 Finally, we would like to mention that in reality, in the case of very 
large $\textit{N}$, the time dependent fluctuations
of
 external sources become important. They are not described by Eq.1 and
 the problem remains unsolved.

This work was
 supported by the Division of Material Sciences,
U.S.National Science Foundation under Contract No. DMR-o228104, and by
grant No. 01-02-17794 of Russian Science Foundation.
We would like to thank L. Levitov for valuable discussions.

\newpage

\begin{figure}
  \centerline{\epsfxsize=10cm \epsfbox{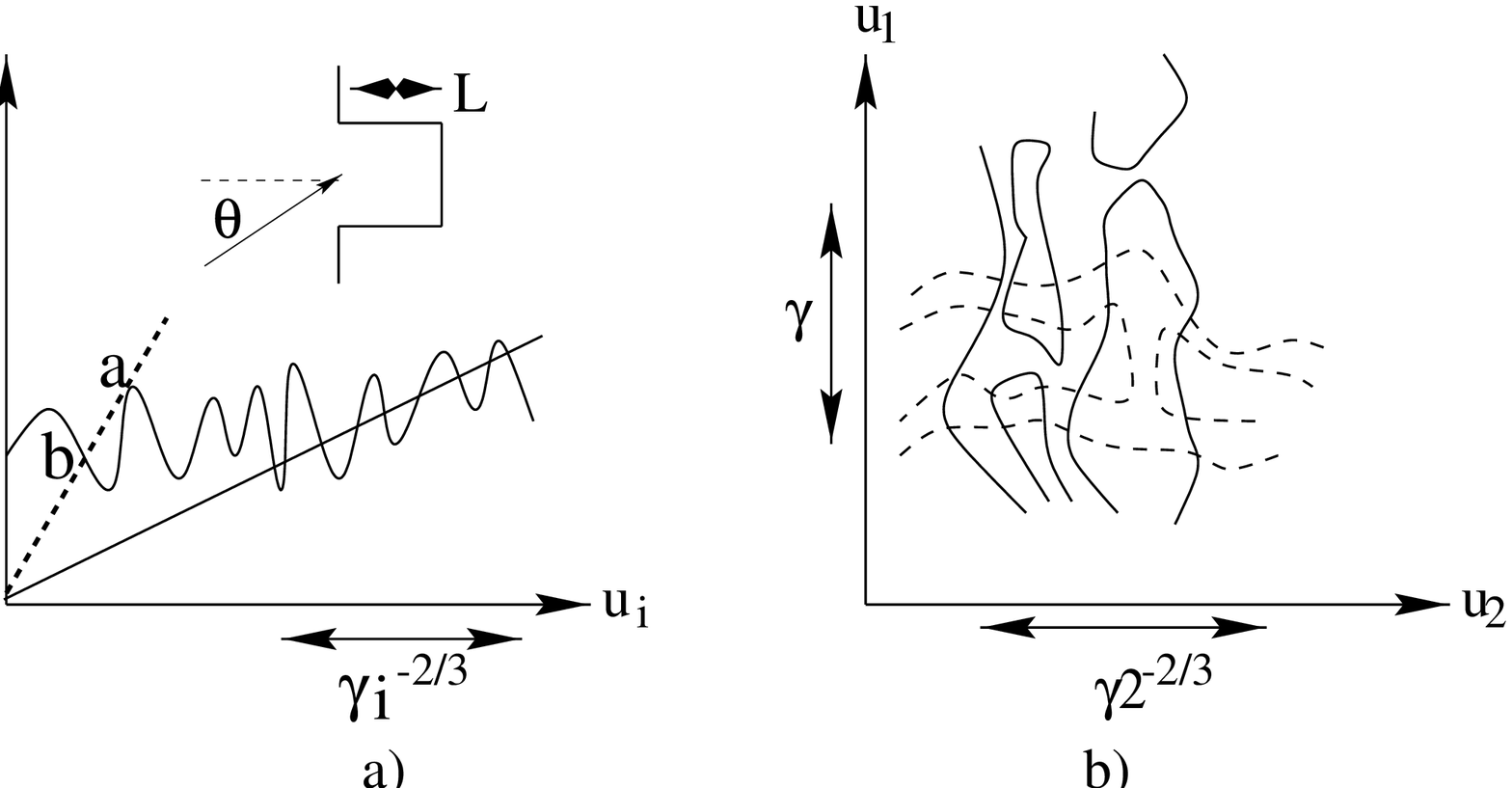}}
  \caption{a) Graphical solution of Eq.7. The wavy line corresponds to
$F_{i}(...\bar{u}_{i}...)$ and  the straight lines corresponds to
$\gamma^{-1}i^{2/3}\bar{u}_{i}$ at different values of $\gamma$.
The dashed line illustrates the critical instability point, when a 
solution
of Eq.7 becomes nonunique.
b) The solid lines correspond to the intersection
 $F_{1}(\bar{u}_{1}, \bar{u}_{2})$ and $\gamma^{-1}\bar{u}_{1}$, while the
dashed
lines correspond to the intersection of 
$F_{2}(\bar{u}_{1},
\bar{u}_{2})$ and
$\gamma^{-1}2^{2/3}\bar{u}_{2}$.} \
  \label{fig:fig1}
\end{figure}

\newpage

\begin{figure}
  \centerline{\epsfxsize=10cm \epsfbox{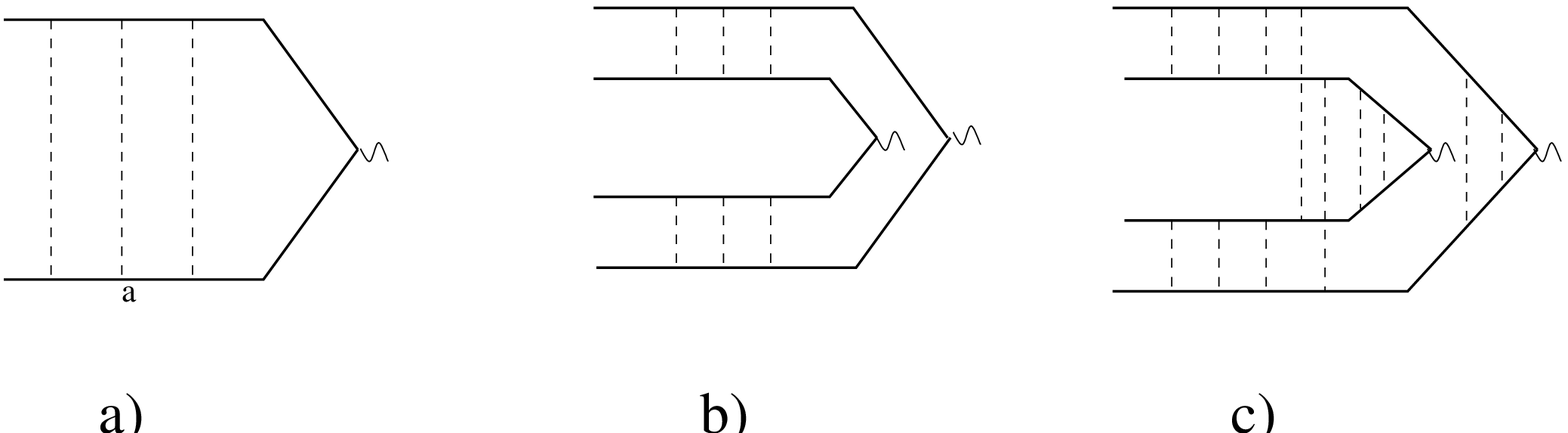}}
  \caption{a) Diagrams describing $\langle n({\bf r})\rangle$. Solid lines correspond to Green functions of 
Eq.1 with $\beta=0$. Dashed lines correspond to 
$\pi \delta({\bf r-r'})/lm^{2}$. b) and c) Diagrams describing Eq.24.
 The inner solid lines describe the Green functions which correspond
 to $\{\bar{u}_{i}\}$, while the outer solid lines correspond to 
$\{\bar{u}_{i}+\Delta \bar{u}_{i}\}$.} \
  \label{fig:fig1}
\end{figure}


\begin{thebibliography}{99}

 
\bibitem{Landau}L. Landau, E. Lifshitz, "Electrodynamics of 
continuous media", 1968.

\bibitem{Zakharov}V.E. Zakharov, V.S. Lvov, G. Falkovich, "Kolmogorov
spectra of
turbulence" Springer-Verlag, 1992.

\bibitem{Kadomzev}B.B. Kadomzev, "Collective phenomena in plasma", 
Nauka, Moscow, 1976.
\bibitem{ZuyzinSpivak}A. Zyuzin, B. Spivak, "Langevin description 
of mesoscopic fluctuations in disordered media",
Sov.Phys.JETP.{\bf 66}, 560-565 ,(1987).
\bibitem{Kane} S. Feng, C. Kane, P.A. Lee, A.D. Stone,
"Correlations and Fluctuations of Coherent Wave Transmission through
Disordered Media", Phys.Rev.Lett. {\bf 61}, 834-837, 1988.
\bibitem{ZyuzinSpivakRev}B. Spivak, A. Zyuzin,
"Mesoscopic Fluctuations of Current Density
in Disordered Conductors" In "Mesoscopic Phenomena in Solids" Ed.
B. Altshuler, P. Lee, R. Webb, Modern Problems in Condensed Matter
Sciences vol.30, 1991.

\bibitem{SpivakZyuzinNL} B. Spivak, A. Zyuzin, "Mesoscopic sensitivity 
of speckles in disordered
nonlinear media to changes of disordered potential"
Phys.Rev. Lett.{\bf 84}, 1970-1073, (2000).

\bibitem{LeeStone}P.A. Lee, A.D. Stone, "Universal Conductance
Fluctuations
in Metals", Phys.Rev.Lett.{\bf 55}, 1622-1625, (1985).

\bibitem{AltshulerSpivak}B. Altshuler, B. Spivak, "Change of random
potential realization and conductivity of
small size sample", JETP Lett.{\bf 42}, 447-449, (1986);
B.D.Simons, B.L.Altshuler, "Universalities in the spectra of disordered
and chaotic systems", Phys.Rev. {\bf B 48}, 5422-5425, (1993).

\bibitem{FengLeeStone}S. Feng, P. Lee, A.D. Stone,, "Sensitivity of the
Conductance of a Disordered Metal to the Motion of a Single Atom" 
Phys.Rev.Lett.{\bf 56}, 1960-1063, (1986).

\bibitem{Abrikosov}A. Abrikosov, L. Gorkov, I. Dzyialoshinski, "Methods of
quantum
field theory in statistical physics" NY,1963.

\bibitem{finkelshtein}, " A.M. Finkelshtein,
"ELectron Liquid in Disordered Conductors", Sov.Sci.Rev. A., Phys. {\bf
14}, 1-42, (1990).
\bibitem{skipetrov1} S.E. Skipetrov, "Temporal fluctuations of waves in 
weakly nonlinear disordered media",  Phys. Rev. {\bf E 63}, 056614(1-15),
(2001).
\bibitem{skipetrov2} S.E. Skipetrov, R. Maynard, "Diffusive waves 
in nonlinear disordered 
media", NATO Science Series, Mathematics, Physics and Chemistry. Vol 107,
Ed. by B.A. van 
Tiggelin and S.E. Skipetrov (Kluwer Academic Publishers, Dordrecht, 2003) 
cond-mat/0303127.
\bibitem{skipetrov3} S.E. Skipetrov, "Langevin description of speckle
 dynamics in nonlinear disordered media", Phys. Rev. {\bf E 67}, 016601,
2003; S.E. Skipetrov, "Instability of speckle patterns in random media
 with noninstantaneous Kerr nonlinearity", Optics Letters, {\bf 28},
646-648, (2003).

\end{thebibliography}
\end{document}